\documentclass[twoside,twocolumn,9pt]{article}
\usepackage{extsizes}
\usepackage[super,sort&compress,comma]{natbib} 
\usepackage[version=3]{mhchem}
\usepackage[left=1.5cm, right=1.5cm, top=1.785cm, bottom=2.0cm]{geometry}
\usepackage{balance}
\usepackage{mathptmx}
\usepackage{sectsty}
\usepackage{graphicx} 
\usepackage{lastpage}
\usepackage[format=plain,justification=justified,singlelinecheck=false,font={stretch=1.125,small,sf},labelfont=bf,labelsep=space]{caption}
\usepackage{float}
\usepackage{fancyhdr}
\usepackage{fnpos}
\usepackage[english]{babel}
\addto{\captionsenglish}{%
  
}
\usepackage{array}
\usepackage{droidsans}
\usepackage{charter}
\usepackage[T1]{fontenc}
\usepackage[usenames,dvipsnames]{xcolor}
\usepackage{setspace}
\usepackage[compact]{titlesec}
\usepackage{hyperref}

\usepackage{epstopdf}

\definecolor{cream}{RGB}{222,217,201}

\usepackage{dcolumn}
\usepackage{bm}
\usepackage[nolist]{acronym}
\usepackage[caption=false]{subfig}
\usepackage{chemformula}
\usepackage[]{siunitx}
\usepackage{tabularx}
\newcolumntype{P}[1]{>{\centering\arraybackslash}p{#1}}
\usepackage{sansmath}
\usepackage[absolute]{textpos}

\begin{document}

\pagestyle{fancy}
\thispagestyle{plain}
\fancypagestyle{plain}{
\renewcommand{\headrulewidth}{0pt}
}

\makeFNbottom
\makeatletter
\renewcommand\LARGE{\@setfontsize\LARGE{15pt}{17}}
\renewcommand\Large{\@setfontsize\Large{12pt}{14}}
\renewcommand\large{\@setfontsize\large{10pt}{12}}
\renewcommand\footnotesize{\@setfontsize\footnotesize{7pt}{10}}
\makeatother

\renewcommand{\thefootnote}{\fnsymbol{footnote}}
\renewcommand\footnoterule{\vspace*{1pt}%
\color{cream}\hrule width 3.5in height 0.4pt \color{black}\vspace*{5pt}} 
\setcounter{secnumdepth}{5}

\makeatletter 
\renewcommand\@biblabel[1]{#1}            
\renewcommand\@makefntext[1]%
{\noindent\makebox[0pt][r]{\@thefnmark\,}#1}
\makeatother 
\renewcommand{\figurename}{\small{Fig.}~}
\sectionfont{\sffamily\Large}
\subsectionfont{\normalsize}
\subsubsectionfont{\bf}
\setstretch{1.125} 
\setlength{\skip\footins}{0.8cm}
\setlength{\footnotesep}{0.25cm}
\setlength{\jot}{10pt}
\titlespacing*{\section}{0pt}{4pt}{4pt}
\titlespacing*{\subsection}{0pt}{15pt}{1pt}

\fancyfoot{}
\fancyfoot[LO,RE]{\vspace{-7.1pt}\includegraphics[height=9pt]{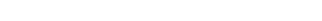}}
\fancyfoot[CO]{\vspace{-7.1pt}\hspace{13.2cm}\includegraphics{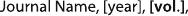}}
\fancyfoot[CE]{\vspace{-7.2pt}\hspace{-14.2cm}\includegraphics{head_foot/RF}}
\fancyfoot[RO]{\footnotesize{\sffamily{1--\pageref{LastPage} ~\textbar  \hspace{2pt}\thepage}}}
\fancyfoot[LE]{\footnotesize{\sffamily{\thepage~\textbar\hspace{3.45cm} 1--\pageref{LastPage}}}}
\fancyhead{}
\renewcommand{\headrulewidth}{0pt} 
\renewcommand{\footrulewidth}{0pt}
\setlength{\arrayrulewidth}{1pt}
\setlength{\columnsep}{6.5mm}
\setlength\bibsep{1pt}

\makeatletter 
\newlength{\figrulesep} 
\setlength{\figrulesep}{0.5\textfloatsep} 

\newcommand{\topfigrule}{\vspace*{-1pt}%
\noindent{\color{cream}\rule[-\figrulesep]{\columnwidth}{1.5pt}} }

\newcommand{\botfigrule}{\vspace*{-2pt}%
\noindent{\color{cream}\rule[\figrulesep]{\columnwidth}{1.5pt}} }

\newcommand{\dblfigrule}{\vspace*{-1pt}%
\noindent{\color{cream}\rule[-\figrulesep]{\textwidth}{1.5pt}} }

\makeatother

\twocolumn[
  \begin{@twocolumnfalse}
{\includegraphics[height=30pt]{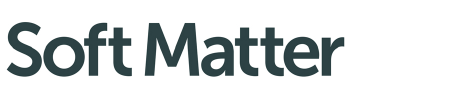}\hfill\raisebox{0pt}[0pt][0pt]
{\includegraphics[height=55pt]{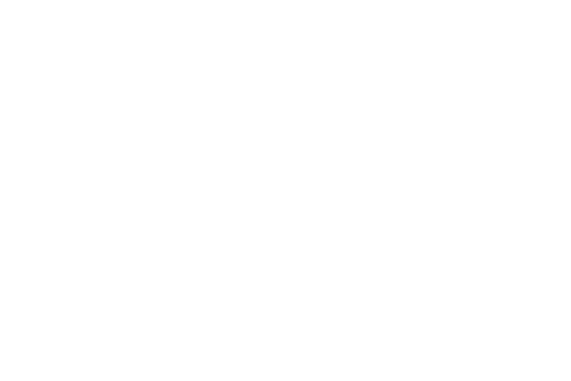}}\\[1ex]
\includegraphics[width=18.5cm]{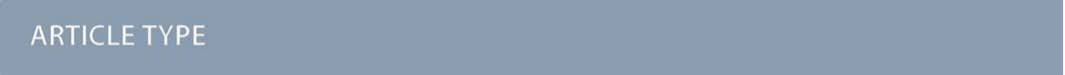}}\par
\vspace{1em}
\sffamily
\begin{textblock*}{200mm}(25mm,10mm)
Accepted Manuscript first published on 20 Feb 2024 in \textit{Soft Matter}, 2024, \textbf{20}, 2831–2839; DOI: \href{https://doi.org/10.1039/D3SM01625K}{10.1039/D3SM01625K}
\end{textblock*}

\begin{tabular}{m{4.5cm} p{13.5cm} }

\includegraphics{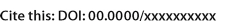} & \noindent\LARGE{\textbf{Rate-independent hysteretic energy dissipation in collagen fibrils\dag}} \\
\vspace{0.3cm} & \vspace{0.3cm} \\

 & \noindent\large{Robert Magerle,\textit{$^{a}$} Paul Zech,\textit{$^{a}$} Martin Dehnert,\textit{$^{a}$} Alexandra Bendixen,\textit{$^{a}$} and Andreas Otto\textit{$^{a}$}} \\

\includegraphics{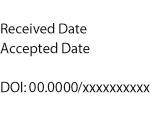} & \noindent\normalsize{
    \begin{sansmath}
    Nanoindentation cycles measured with an atomic force microscope on hydrated collagen fibrils exhibit a rate-independent hysteresis with return point memory.
    This previously unknown energy dissipation mechanism describes in unified form elastoplastic indentation, capillary adhesion, and surface leveling at indentation velocities smaller than $\qty{1}{\micro\metre\per\s}$, where viscous friction is negligible. 
    A generic hysteresis model, based on force-distance data measured during one large approach-retract cycle, predicts the force (output) and the dissipated energy for arbitrary indentation trajectories (input). 
    While both quantities are rate independent, they do depend nonlinearly on indentation history and on indentation amplitude.  
    \end{sansmath}
} \\

\end{tabular}

 \end{@twocolumnfalse} \vspace{0.6cm}

  ]

\renewcommand*\rmdefault{bch}\normalfont\upshape
\rmfamily
\section*{}
\vspace{-1cm}

\footnotetext{\textit{$^{a}$~Fakult{\"a}t f{\"u}r Naturwissenschaften, Technische Universit{\"a}t Chemnitz, 09107 Chemnitz, Germany.}}

\footnotetext{\dag~Electronic Supplementary Information (ESI) available: Experimental protocol for cyclic FD measurements, additional FD data illustrating steady-state hysteresis, and dissipated energy measured in cyclic indentation experiments. See DOI: 10.1039/cXsm00000x/}



\section{Introduction}
\begin{figure}[h!]
    \centering
    \includegraphics[width=\columnwidth]{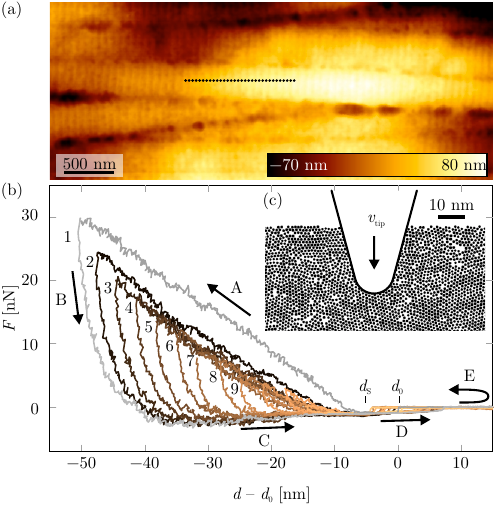}
    \caption{
    \begin{sansmath}
    Steady-state hysteresis in cyclic AFM nanoindentation measurements. (a) AFM height image of native hydrated collagen fibrils. 
    (b) Cyclic measurements of the force $F$ as a function of the tip--sample distance $d-d_0$ taken at an overlap position show steady-state hysteresis after the initial conditioning cycle (gray). Position $d_0$ is the onset of the capillary force, indents are enumerated. The indent is made in the radial direction, perpendicular to the fibril axis, with $v_{\text{tip}} = \SI{0.05}{\micro\metre\per\second}$. The measurements are made on the collagen fibril marked in (a). 
    (c) Schematic cross-section perpendicular to the fibril axis at the tip apex ($\qty{8}{\nano\m}$ radius, $\ang{30}$ opening angle). Dots represent tropocollagen molecules.          
    \end{sansmath}
    }
    \label{fig:forc}
\end{figure}

Collagen fibrils are the tensile-load bearing components in connective tissue such as skin, ligament, tendon, cartilage, and bone.\cite{Fratzl2008} 
A tendon owes its remarkable tensile strength and bending flexibility to the molecular structure of type I collagen and the supramolecular order in collagen fibrils.\cite{Gautieri2011} 
Type I collagen forms $\qty{300}{\nano\m}$ long triple-helical protein assemblies (tropocollagens), which are staggered and aligned along the fibril's axis.\cite{Orgel2006} This results in a smectic-like order with $\qty{67}{\nano\m}$ repeat distance. This periodic D-band pattern is visible in electron microscopy \cite{Hukins1977,Hulmes1981} and \ac{AFM} images (Fig.\,\ref{fig:forc}a). 
In the plane perpendicular to the fibril axis, the tropocollagen order is best described as being of polycrystalline texture, comprising small crystalline domains ordered quasi-hexagonally.\cite{Hulmes1981,Hulmes1995} 

Finding out how collagen molecules and fibrils respond to mechanical stress and how energy is dissipated are topics of current research, whose purpose it is to reveal the molecular mechanisms involved \cite{Gautieri2011,Masic2015,Andriotis2018,Zapp2020} and to provide an understanding of the changes connective tissues undergo due to ageing and disease.\cite{Stolz2009,Kemp2012,Gonzalez2014,Chi2022}
To achieve this, \ac{AFM} is used to image and measure mechanical properties of individual collagen fibrils in the hydrated state, either in buffer solution\cite{Stolz2004,Stolz2009,Grant2009,Baldwin2014,Andriotis2014,Andriotis2018,Andriotis2019,Andriotis2023} or in humid air.\cite{Spitzner2015,Uhlig2017,Magerle2020}
The hysteresis between \ac{FD} data measured during tip approach and tip retraction (Fig.\,\ref{fig:forc}b) is commonly referred to as viscoelastic behavior,\cite{Grant2009,Uhlig2017} since it is similar to the behavior of much softer and more fluid-like biological matter.\cite{Fung1993,Efremov2020} 
This implies that viscous friction is the main cause of energy dissipation and leads to the widespread assumption that the friction force is proportional to tip velocity $v_{\text{tip}}$.\cite{Garcia2006,Proksch2012,Efremov2020} 

\begin{figure}[t!]
    \centering
    \includegraphics[width=\columnwidth]{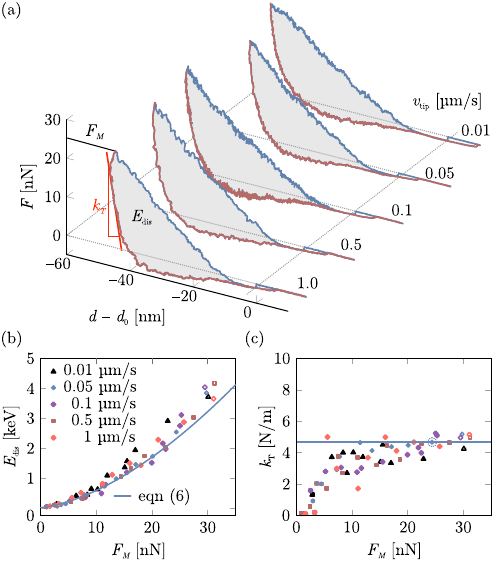}
    \caption{
    \begin{sansmath}
    Rate-independent energy dissipation. (a) \ac{FD} data measured with different tip velocities $v_\text{tip}$ during tip approach (blue) and tip retraction (red). The measurements are made on the collagen fibril marked in Fig.\,\ref{fig:forc}a.
    (b) $E_{\text{dis}}$ as a function of $F_M$ measured with different $v_\text{tip}$ in cyclic indentation experiments. 
Data points with open symbols are from initial conditioning cycles. The curve is the prediction with the rate-independent hysteresis model for $v_\text{tip} = \SI{0.05}{\micro\metre\per\second}$. (c) $k_T$ as a function of $F_M$. The line and the circle indicate the value used in the hysteresis model's prediction shown in (b). 
    \end{sansmath}
    } 
    \label{fig:Edis_experiment}
\end{figure}

Here, we report our finding that energy dissipation does in fact not depend on the indentation rate for  $v_\text{tip} \le \SI{1}{\micro\metre\per\second}$ (Fig.\,\ref{fig:Edis_experiment}).
The nanoindentation response of hydrated collagen fibrils displays a rate-independent hysteresis with return point memory that depends only on the last return point. 
We find that the return points are the tip positions where the tip motion changes its direction. 
This energy dissipation mechanism dominates at slow deformation rates and cannot be captured with linear and quasilinear viscoelastic mechanical models \cite{Fung1993} that are based on combinations of spring and dashpot elements.\cite{Dufrene2013,Krieg2019,Efremov2020} 

Return point memory is a specific type of memory formation in hysteretic systems that allows the system to return to a previous state---the return point---at a later time.\cite{Sethna1993,BERTOTTI1998} 
After a return point has been revisited and passed through again, the memory of that return point is wiped out.
This is observed in ferromagnetic materials \cite{Preisach1935,Sethna1993} and in thermoelastic martensitic transformations of shape memory alloys.\cite{Delaey1974,Ortin1992}
For further examples see ref.\,\citenum{Sethna1993} and references therein. 
Return point memory is also found in the elastoplastic deformation of cast iron \cite{Prandtl1928} and in cyclic nanoindentation of Fe-based amorphous alloys.\cite{Lashgari2018} 
The analogy with the latter shows that hydrated type I collagen fibrils are soft, ductile solids with elastoplastic deformation behavior. This is in accordance with the finding that native type I collagen fibrils are in the glassy state at physiological conditions.\cite{Gevorkian2011} 
Collagen fibrils, however, are soft enough that the surface tension causes a capillary bridge to form  and the surface to level out after the AFM tip is withdrawn.

What is more, we present a generic hysteresis model that describes in unified form sequences of the aforementioned distinct mechanical phenomena occurring in AFM-based nanoindentation experiments: elastoplastic indentation, capillary adhesion, and surface leveling (EPICAL). 
The model is based on measured \ac{FD} data and it predicts the force (output) for arbitrary indentation trajectories (input) over a wide range of tip velocities and indentation amplitudes. 
This accurately predicts the dissipated energy. 
The model also accounts for attractive forces due to adhesion and the capillary interaction, which are known hysteretic energy dissipation mechanisms in \ac{AFM}.\cite{Garcia2012,Zitzler2002} 

\section{Results and disscusion}
\subsection{AFM-based nanoindentation experiments}
We study native collagen fibrils extracted from chicken tendon, deposited onto a polished Si wafer, and measured in humid air at $\SI{90}{\percent}$ relative humidity (Fig.\,\ref{fig:forc}a). 
For details on materials and methods, see Appendix A.
For each tip velocity, we measure cyclic \ac{FD} data at increments of $\qty{32}{\nano\m}$ along the crest of a single collagen fibril (indicated by black dots in Fig.\,\ref{fig:forc}a). This allows us to distinguish between the mechanical response of overlap regions and that of gap regions. 
We measure \ac{FD} data by approaching the tip with a velocity $v_\text{tip}$ until the maximal-force setpoint $F_{M}$ is reached. 
From this point on, the tip is made to retract with the same velocity. 
Remaining in the same position, this is followed up by repeated indentation cycles with an incrementally decreasing setpoint $F_M$ (Fig.\,\ref{fig:forc}b). The temporal evolution of the corresponding force probing protocol is shown in Fig.\,S1\dag.

\subsection{Steady-state hysteresis}
In the following, we focus on the overlap regions. Gap regions display the same type of hysteresis behavior (data not shown) but are about 
15\% softer.\cite{Baldwin2014,Magerle2020} The shape of the repulsive section of FD data (Figs.\,\ref{fig:forc}b and \ref{fig:Edis_experiment}a) is typical for elastoplastic deformation, much like that occurring in nanoindentation experiments on glassy polymers,\cite{Briscoe1998} inorganic solids,\cite{Oliver1992} and metallic glasses\cite{Lashgari2018}. 
On hydrated collagen fibrils, however, the remnant indentation is negligible ($<\qty{1}{\nm}$).
This counterintuitive effect is caused by surface tension, which we outline in greater detail. 
The first indentation--retraction cycle (gray line in Fig.\,\ref{fig:forc}b) initializes the tip--sample contact. 
All following indentation--retraction cycles share a common approach trajectory and a common retraction trajectory. This indicates steady-state hysteresis. Further evidence is provided by cyclic nanoindentation measurements performed with a constant maximal-force setpoint $F_{M}$ that perfectly overlap (Fig.\,S2\dag). 

During hysteresis, the following sequence of distinct mechanical processes occurs, which we label with capital letters as shown in Fig.\,\ref{fig:forc}b: 
(A) Elastoplastic deformation during tip indentation. 
(B) Upon tip retraction, the force decreases due to the fibril's elastic response. 
(C) This is followed by a regime of attractive forces, caused by adhesion, by the capillary force, and by the formation of a capillary bridge. 
(D) Retracting the tip well beyond the original surface breaks the capillary bridge at $d_\text{off}$.
(E) During further tip retraction and the following tip-approach period, the surface tension causes surface leveling.
As a result, the onset of the capillary force occurs approximately at the same position $d_0$ for all indentation cycles.
The overall behavior can be characterized as being shape reversible while being dissipative at the same time.

\subsection{Energy dissipation}
The shape of \ac{FD} curves is independent of tip velocity in the range from $\num{0.01}$ to $\qty{1}{\micro\metre\per\s}$, as shown for \ac{FD} data sets measured after the initial conditioning cycle (Fig.\,\ref{fig:Edis_experiment}a). 
The energy dissipated during each approach-retract cycle, $E_\text{dis}$, is represented by the area enclosed between the \ac{FD} data measured during tip approach and tip retraction (Fig.\,\ref{fig:Edis_experiment}). 
By integrating we obtain $E_\text{dis}$ as a function of the maximal-force setpoint $F_M$ for every \ac{FD} cycle. 
The dissipated energy $E_\text{dis}$ is independent of $v_\text{tip}$ and depends only on the maximal-force setpoint $F_M$ (Fig.\,\ref{fig:Edis_experiment}b). 
Figure\,\ref{fig:Edis_experiment}b shows the data from cyclic nanoindentation experiments for each tip velocity. 
For each velocity, we repeated this measurement at multiple locations along the same collagen fibril. 
All $E_\text{dis}$ data measured at the overlap regions are shown in Fig.\,S3$\dag$ and display the same behavior as in Fig.\,\ref{fig:Edis_experiment}b. 
For each tip velocity and $F_M$ value, the scatter (standard deviation) of the $E_\text{dis}$ data is typically below 20\% (Fig.\,S3$\dag$). 

\subsection{Contact stiffness}
We interpret the force decrease during a small tip retraction as the fibril's elastic response. This is evident in experiments with multiple small retract-approach cycles that are discussed in Section \ref{sec:rpm}.
We determine the contact stiffness $k_T$ from the slope of \ac{FD} data measured during tip retraction (Fig.\,\ref{fig:Edis_experiment}a), which is comparable to the Oliver-Pharr method.\cite{Oliver1992} 
The computational procedure is detailed in Section \ref{sec:comparison}. We find that the contact stiffness $k_T$ is also independent of tip velocity. It increases only slightly for $F_M > \SI{5}{\nano\newton}$ (Fig.\,\ref{fig:Edis_experiment}c).
In the hysteresis model, we use the $k_T$ value obtained for the second indent measured with $v_{\text{tip}} = \SI{0.05}{\micro\metre\per\second}$, that represents the steady-state situation.

\subsection{Repulsive forces}\label{sec:repforce}
In our hysteresis model, we describe the repulsive forces phenomenologically with an algebraic function resembling the functional form of common contact models; see below, eqn\,(\ref{eq:fA_fR}a). The Hertz model as well as related contact models \cite{Cappella1999} are only valid for elastic materials and cannot describe plastic deformation. Despite this limitation, the functional form of elastic models is frequently used for describing \ac{FD} data beyond the elastic regime. \cite{Efremov2020,Dufrene2013,Krieg2019, Uhlig2017, Magerle2020, Dehnert2018}

\begin{figure*}[t!]
    \centering
    \includegraphics[width=\textwidth]{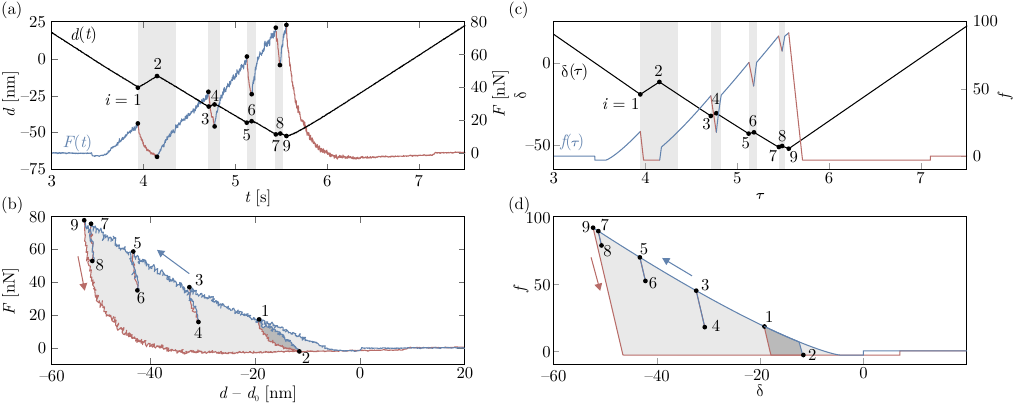}
     \caption{
     \begin{sansmath}
     Return point memory in indentation trajectories with multiple retract-approach cycles. Return points ($\bullet$) are sequentially numbered. (a) Tip--sample distance $d(t)$ and force $F(t)$ measured between the \ac{AFM} tip and a collagen fibril in humid air during an indentation trajectory involving several approach (blue) and retract periods (red) with $v_\text{tip} = \SI{0.04}{\micro\metre\per\second}$. This measurement is made on an isolated collagen fibril deposited on a Si substrate.
(b) Same $F$ data plotted as a function of tip--sample distance $d-d_0$ which takes the cantilever bending into account. (c) Input $\delta(\tau)$ (thin line) determined from the measured indentation trajectory shown in (a) and the output $f(\tau)$ (thick line) calculated with the hysteresis model. (d) The same data plotted in the $\delta$-$f$ plane.
\end{sansmath}
}
    \label{fig:play_experiment}
\end{figure*}

\subsection{Return point memory}\label{sec:rpm}
Experiments with more complex tip trajectories, including multiple intermediate retract-approach cycles (gray regions in Fig.\,\ref{fig:play_experiment}a), clearly show rate-independent hysteretic behavior with return point memory. 
At the end of each intermediate retract-approach cycle the system returns to the starting point of each cycle (1$\rightarrow$2$\rightarrow$1, 3$\rightarrow$4$\rightarrow$3, 5$\rightarrow$6$\rightarrow$5, 7$\rightarrow$8$\rightarrow$7 in Fig.\,\ref{fig:play_experiment}b). This is a clear signature of the return point memory.
Since no inner loops occur, the system state depends only on the last return point (Fig.\,\ref{fig:play_experiment}a,b).

In our case, the return points are the positions $d_i^*$ where the tip velocity changes its sign along with corresponding times $t_i^*$ and $i=1,\dots,N$ (Fig.\,\ref{fig:play_experiment}).
For a starting point well above the surface, an \ac{FD} trajectory is uniquely determined by its sequence of return points, which carry the system's built-in memory.
In Fig.\,\ref{fig:play_experiment}, for instance, the return point $i = 3$ marks the start of an intermediate retract-approach cycle.
After passing this point, the system enters an excursion, whereby the force starts to decrease linearly.
During these intermediate excursions, the force changes approximately linearly (elastic). 
It has a contact stiffness similar to $k_T$ (Fig.\,\ref{fig:Edis_experiment}c).
Before the return point $i = 3$, and after reaching this position again, the force increases with a power of $d$ following the outer hysteresis loop.
This corresponds to the \ac{FD} data measured with a large indentation amplitude (Fig.\,\ref{fig:Edis_experiment}).

\subsection{Rate-dependent phenomena}
Hysteresis is also observed for $v_\text{tip} > \SI{1}{\micro\metre\per\second}$, at which point a small phase lag between force (stress) and indentation (strain) causes the formation of inner hysteresis loops due to viscous friction (see Appendix B). In the studied velocity range, stress relaxation and creep progress more slowly than the tip velocity, which is why they do not influence the time evolution of $F(t)$ occurring during the measurement of \ac{FD} data (see Appendix C).

\subsection{Hysteresis model}
Our aim is to produce a concise model that encompasses the following key features of the tip--sample interaction for arbitrary tip indentation trajectories: the repulsive force during initial tip indentation, the contact stiffness during intermediate retract-approach cycles, and the attractive force caused by adhesion and capillary bridge formation, followed by surface leveling. 

The first step towards our goal is to present a hysteresis model with return point memory for the domain $d \le d_{0}$.
Then we add the elementary hysteresis for capillary bridge formation and its collapse in the domain $d > d_{0}$.
For modeling, we introduce the dimensionless quantities force $ f = F/\SI{1}{\nano\newton} $, distance $ \delta = (d-d_0)/\SI{1}{\nano\metre} $, and time $ \tau = t/\SI{1}{\second} $.

For $\delta \le 0$, we deduce a phenomenological hysteresis model based on the form of measured \ac{FD} data by defining three time-independent functions that describe three characteristic regimes of the tip-sample interaction: $f_A(\delta)$ for the tip approach trajectory, $f_T(\delta,\delta_i^*)$ for the elastic response during intermediate retract-approach cycles, and $f_R(\delta)$ for the attractive-force limit of the retract trajectories:
\begin{subequations}
\begin{align}
    f_A(\delta) &= \begin{cases}
                f_c, & \delta_s < \delta \\
                \kappa_A (\delta_s - \delta)^{\gamma} + f_c, & \delta \le \delta_s ,
            \end{cases} \\
    f_T(\delta,\delta_i^*) &= f(\delta_i^*)-\kappa_T (\delta-\delta_i^*) ,\\        
    f_R(\delta) &= f_c.        
\end{align}
\label{eq:fA_fR}
\end{subequations}
Here, $\delta=\delta(\tau)$ is the actual input of the system at time $\tau$, $\delta_i^* = \delta(\tau^*_i) $ is the last return point, $f_c$ is the attractive (capillary) force, and $\delta_s$ the surface position. 
The function $f_A(\delta)$ resembles the functional form of common contact models.\cite{Cappella1999} The parameter $\kappa_T = k_T/\SI[per-mode = power]{1}{\newton\per\metre}$ is the dimensionless contact stiffness.
The stiffness $\kappa_A$ is determined by the tip's shape (radius and opening angle) and the sample's yield strength.

\begin{figure}[t!]
    \centering
    \includegraphics[width=\columnwidth]{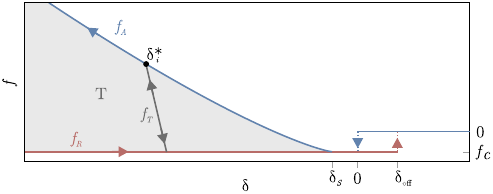}
    \caption{
    \begin{sansmath}
    Input-output diagram of the EPICAL hysteresis model.
The multistable region T is enclosed by the functions $f_A(\delta)$ and $f_R(\delta)$ for $\delta \le 0$.  
The function $f_T(\delta,\delta_i^*)$ describes the elastic response within the region T.
The elementary hysteresis due to capillary bridge formation occurs for $0 < \delta < \delta_\text{off}$.
\end{sansmath}
}
    \label{fig:model_play}
\end{figure}

This describes the input-output diagram of the hysteresis model (Fig.\,\ref{fig:model_play}). 
For $\delta < \delta_s$, and a given input $\delta$, multiple output values $f$ are possible within the transition region T; the trajectory taken is determined by the system's entry point into region T, that are return points of the input $\delta$. Complex tip trajectories include multiple return points $\delta_i^*$, which represent the system's memory. 
This memory is implemented by defining the output trajectory $f(\tau)$ piecewise between return points $\delta_i^*$, which are the points where the tip velocity $\mathrm{d}\delta / \mathrm{d}\tau$ changes its sign along with the corresponding times $\tau^*_i$ and $i=1,\dots,N$:      
\begin{equation}
    f(\tau) = \begin{cases}
                \text{min}\{f_T(\delta,\delta_i^*),f_A(\delta)\},\quad \delta-\delta_i^* < 0 \\
                \text{max}\{f_T(\delta,\delta_i^*),f_R(\delta)\},\quad \delta-\delta_i^* > 0, \\
                \end{cases}
\label{eq:output_1}                 
\end{equation}
where $\delta=\delta(\tau)$, $\delta_i^* = \delta(\tau^*_i) $, and $\tau^*_i < \tau \le \tau^*_{i+1}$. An equivalent, but more compact form of eqn\,(\ref{eq:output_1}) is
\begin{equation}
    \mathrm{\hat \Gamma}[\delta](\tau) := f(\tau) = \text{max} \Bigl\{ f_R(\delta) , \text{min} \bigl\{ f_A(\delta) , f_T(\delta,\delta_i^*) \bigl\} \Bigl\} .
    \label{eq:output}  
\end{equation}

Equation\,(\ref{eq:output}) defines a hysteresis operator $ \mathrm{\hat \Gamma}[\delta] $ that acts on a whole input trajectory $\delta(\tau)$ and yields the output $ f(\tau) = \mathrm{\hat \Gamma}[\delta](\tau) $.
For numerical computations eqn\,(\ref{eq:output_1}) can be understood as an iterative rule that determines the output $f(\tau)$ at time $\tau$ for an infinitesimally small indentation step $ \Delta\delta = \delta(\tau)-\delta(\tau') $ from a given pair $(\delta(\tau'),f(\tau'))$ at time $\tau'$.
It should be noted that this rule, and the corresponding hysteresis operator $ \mathrm{\hat \Gamma}[\delta] $, are rate-independent given that eqns\,(\ref{eq:output_1}) and (\ref{eq:output}) do not depend on the time step $\Delta\tau = \tau-\tau'$.

The hysteresis model displays multistability for $\delta < \delta_s$ since, for a given input  $\delta$, a range of output values $f$ are possible and the model displays return point memory only on the previous return point from the sequence of return points $\delta^*_{i}$.
In the region $0 < \delta < \delta_\text{off}$, capillary bridge formation, and its collapse, cause bistability which is described with an elementary hysteresis model.\cite{Mayergoyz1991a}
For $ \delta(\tau) \ge \delta_\text{off} $, the output is unique with $ f(\tau) = 0 $.
In this regime, surface tension causes the leveling of the surface, and with it the system loses its memory.

The input--output behavior is summarized in Fig.\,\ref{fig:model_play}.
What is more, the hysteresis operator $ \mathrm{\hat \Gamma} $ given by eqn\,(\ref{eq:output}) shows similarity to the generalized stop operator,\cite{bobbio1997,wang2021} which is a generalization of Prandtl's model of elasto-plasticity.\cite{Prandtl1928,Visintin1994,brokate1996,kuhnen2003} The generalized stop operator limits the output to a finite range of values. In our case, these limits are (1) the function $f_A$, which describes the maximal repulsive force due to the material's yield strength, and (2) the function $f_R$, which describes the attractive force due to adhesion and the capillary force.

\subsection{Comparison with experiments}\label{sec:comparison}
The EPICAL hysteresis model highly accurately predicts the force and the dissipated energy for complex tip trajectories.
For a given \ac{FD} data set, measured with large indentation amplitude (as in Fig.\,\ref{fig:forc}b), we determine the parameters that describe the EPICAL hysteresis model by fitting functions corresponding to eqns\,(\ref{eq:fA_fR}a), (\ref{eq:fA_fR}b), and (\ref{eq:output}) to the measured data. To this end, we use MATLAB with the EzyFit 2.44 Toolbox \cite{ezyFit2023} and proceed successively as follows.

\begin{figure}[t!]
    \centering
    \includegraphics[width=\columnwidth]{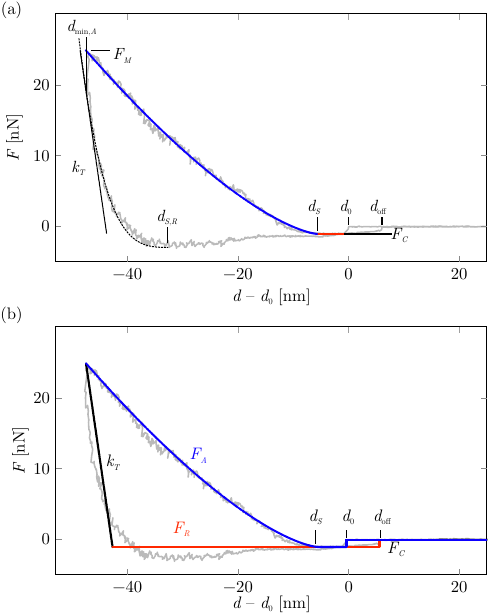}
    \caption{
    \begin{sansmath}
    (a) Piece-wise fitting procedure for obtaining the parameters describing the hysteresis operator. The FD data are measured on a collagen fibril within a bundle of many fibrils. (b) Force predicted with the hysteresis operator for the entire indentation cycle.
     \end{sansmath}
    }
    \label{fig:fit}
\end{figure}

1: For every FD curve, the baseline was adjusted with a sine fit to correct the deflection offset and the laser interference effects during the measurement. 2: The position $d_0$, the onset of attractive forces, was determined as described in ref. \citenum{Dehnert2018} (Fig.\,\ref{fig:fit}a). 3: The capillary force $F_C$ is set to the minimal force value within an FD data set measured during tip approach (Fig.\,\ref{fig:fit}a). The corresponding $d$ position was used as a starting point for $d_S$ in the following fitting step. 4: A power-law function $F_A(d)$ with offset $F_C$, which is analogous to eqn\,(\ref{eq:fA_fR}a), was fitted to the FD data measured during tip approach (blue curve in Fig.\,\ref{fig:fit}a). The maximum indentation $d_{\text{min},A}$ and $F_C$ were kept fixed, whereas $\kappa_A$, $\gamma$ and $d_S$ were varied. Note, that the obtained parameters $\kappa_A$ and $\gamma$ are sensitively dependent on $d_S$. This is caused by the scale invariance of the power-law. It still remains that the resulting fitting function $F_A(d)$ is a good approximate description of the particular FD data set with an analytical function. 5: A power-law function 
\begin{equation}
    F_R(d) = k_R \left(d_{S,R}-d\right)^\beta+ F(d_{S,R})  
\label{eq:kR}
\end{equation}
that is analogous to $F_A(d)$ was fitted to the FD data measured during tip retraction (dashed curve in Fig.\,\ref{fig:fit}a) within the fitting range $d_{\text{min},R} < d < d_{S,R}$, where $d_{S,R}$ is the position of the smallest force value;  $k_\text{R}$, $d_{S,R}$, and $\beta$ were fitting parameters, while $F(d_{S,R})$ is kept fixed. From the resulting fit function, the contact stiffness $k_T$ is obtained as the absolute value of the slope at $d_{\text{min},A}$
\begin{equation}
    k_T = \left| -k_R\beta\left(d_{\text{S},R}-d_{\text{min},A}\right)^{\beta-1} \right|.        
\label{eq:kT}
\end{equation}
The dimensionless contact stiffness $\kappa_T = k_T/\SI[per-mode = power]{1}{\newton\per\metre}$.

Fitting parameters for selected data sets are shown in Table \ref{tab:table_fit_parameters}, which allow for the construction of the hysteresis operator according to eqns \ref{eq:fA_fR} and \ref{eq:output}. An example for the force predicted with the hysteresis operator for the entire indentation cycle is shown in Fig.\,\ref{fig:fit}b.

\begin{table}[t!]
\caption{\label{tab:table_fit_parameters}
Fit parameters in dimensionless units. 
}
\begin{tabular}{llll} \hline
FD data set &\text{Fig.\,\ref{fig:forc}b}$^a$ &\text{Fig.\,\ref{fig:forc}a}$^b$ &\text{Fig.\,\ref{fig:play_experiment}}\\ 
AFM cantilever& \text{Type II} & \text{Type II} & \text{Type I}\\
$v_\text{tip}$ & $\SI{0.05}{\micro\metre\per\second}$ & $\SI{0.05}{\micro\metre\per\second}$ & $\SI{0.04}{\micro\metre\per\second}$\\ \hline
 parameter & \text{value} & \text{mean}$\pm$\text{SD} & \text{value}\\
\hline
       $f_c$ & $-1.08$ & $-0.976\pm0.161$ & $-4.77$ \\
       $\delta_s$ & $-10.0$ & $-6.56\pm2.51$ & $-4.62$ \\
       $\kappa_A$ & $0.192$ & 0.116$\pm$0.083$^c$ & 0.728 \\
       $\gamma$ & $1.26$ & $1.52\pm0.211^c$ & $1.25$ \\
       $\kappa_T$ & $3.58$ & $5.32\pm0.589$ & $16.4$ \\
       $f_M$ & $22.8$ & $25.0\pm2.88$ & $76.5$\\ 
\end{tabular}
$^a$Second indentation cycle of FD data shown in Fig.\,\ref{fig:forc}b. $^b$Set of second indentation cycles measured at 15 overlap positions along the line marked in Fig.\,\ref{fig:forc}a. $^c$Note, that the fitting parameters $\kappa_A$ and $\gamma$ are sensitively dependent on $d_S$. This causes the large SD.
\end{table}

The obtained parameters $d_0$, $d_\text{off}$, and $f_c$ describe the capillary interaction, $d_S$, $\kappa_A$, and $\gamma$ describe the elastoplastic indentation, and the contact stiffness $\kappa_T$ describes the elastic response.
For simplicity, we describe adhesion with $f_c$.
This defines the hysteresis model for the particular \ac{FD} data set making it possible to determine the output $f(\tau)$ for any other $\delta(\tau)$ trajectory, for example, the one shown in Fig.\,\ref{fig:play_experiment}a.
Proceeding on this basis, we determine the sequence of return points $\delta^*_{i}$ from the measured trajectory $d(t)$, which yields the $\delta(\tau)$ trajectory shown in Fig.\,\ref{fig:play_experiment}c.
The resulting $f(\delta(\tau))$ trajectory (Fig.\,\ref{fig:play_experiment}c,d) represents all essential features of the measured force trajectory $F(d)$.
Differences are only observed for the first small retract-approach loop starting at $t_1^*$ and the last retract period starting at $t_9^*$, since we assumed trajectories in region T to be linear functions.

\subsection{Dissipated energy}
The dissipated energy is also correctly predicted by the hysteresis model. It is given by the integral $\oint f(\delta)\,\text{d}\delta$ over an indentation cycle. For the indentation cycles with varying $F_M$ (Fig.\,\ref{fig:Edis_experiment}b) we obtain in reduced units:
\begin{equation}
    E_\text{dis}(f_M) = \frac{\left( f_M-f_c \right)^{\frac{1}{\gamma}+1}}{(1+\gamma)\kappa_A^{1/\gamma}} - \frac{\left(f_M-f_c \right)^2}{2 \kappa_T},
\label{eq:Edis}
\end{equation} 
where $f_M$ corresponds to $F_M$. 
The first term is the energy invested during tip indentation and the second is the elastic energy regained during tip retraction. 
The resulting $E_\text{dis}(F_M)$ (solid line in Fig.\,\ref{fig:Edis_experiment}b) does not depend on $v_\text{tip}$ and it describes the measured data very accurately for the entire range of $F_M$. It should be noted, that eqn\,(\ref{eq:Edis}) is the result of a specific set of indentation cycles. 
It slightly underestimates the dissipated energy since, for simplicity's sake, the adhesion is being described with $f_c$.

In general, the energy dissipated during an indentation cycle depends not only on the actual indentation $\delta$ and the indentation amplitude, but also on the last return point. 
Put simply: for small retract-approach amplitudes, the response is elastic, whereas for larger amplitudes, energy is dissipated, provided $f(\tau)$ switches between the boundaries $f_R$ (maximal tensile strength) and $f_A$ (maximal compressive strength).
The first small loop starting at $t_1^*$ (Fig.\,\ref{fig:play_experiment}) exemplifies this. 

\subsection{Discussion}
Collagen fibrils have an anisotropic molecular structure and therefore anisotropic mechanical properties. 
The stress--strain behavior of single collagen fibrils when stretched along the fibril axis is investigated using special micro-electromechanical systems\cite{ShenMEMS2010,ShenMEMS2011,Liu2018} or by attaching single collagen fibrils to AFM cantilevers.\cite{Rijt2006,Yang2008,SvenssonFibril2010} 
Under cyclic loading, partial plastic deformation, steady-state hysteresis, and related energy dissipation are observed. 
These measurements primarily study the deformation and sliding of adjacent tropocollagen molecules along their longitudinal axis. 
Uniaxial stretching along the fibril axis is also the subject of recent molecular dynamics computer simulations.\cite{Gautieri2011,Masic2015,Zapp2020,Suhail2023} 
A specific question in these studies is how different types of crosslinks between adjacent tropocollagen molecules affect the mechanical behavior of collagen fibrils.

In AFM-based nanoindentation experiments, on the other hand, tropocollagen molecules are pushed apart perpendicular to their longitudinal axis and sheared against each other in the plane perpendicular to the fibril axis, as suggested by large-scale molecular dynamics simulations of the nanoindentation response of amorphous\cite{Lashgari2018} and nanocrystalline metals.\cite{Li2020} 
We expect similar processes to occur within the nanocrystalline domain structure of collagen fibrils. 
In addition, the tropocollagen helix structure may deform, as suggested by molecular dynamics simulations of tensile loading of collagen fibrils.\cite{Masic2015}  
The cohesive force between adjacent tropocollagen molecules plays an important role in both fundamental deformation modes, stretching along the fibril axis and shearing perpendicular to the fibril axis, but the two deformation modes are fundamentally different, so nanoindentation experiments cannot be directly compared to stretching along the fibril axis.

\section{Conclusions}
In \ac{AFM} nanoindentation experiments on hydrated collagen fibrils, at slow deformation rates, energy is dissipated via a rate-independent hysteretic process with return point memory and negligible viscous friction. 
The fibril's shape-reversible yet dissipative response is caused by a combination of interactions between elastoplastic indentation, adhesion, capillary force, and surface tension, which all together cause surface leveling. 

Native, hydrated type I collagen fibrils are in the glassy state.\cite{Gevorkian2011} At ambient temperature, they are above the glass-to-rubber transition\cite{Batzer1981, Gevorkian2011, Baldwin2014} (in the leathery regime\cite{Tobolsky1956}). Indeed, glassy polymers\cite{Briscoe1998} as well as inorganic\cite{Oliver1992} and metallic glasses\cite{Lashgari2018} show elastoplastic behavior under cyclic loading and in nanoindentation experiments. The underlying molecular mechanisms are material specific, and a range of models have been developed to describe the elastoplastic behavior of amorphous materials,\cite{Nicolas2018} including different types of soft matter. Therefore, we anticipate that the slow (glassy) molecular dynamics cause the observed hysteretic behavior, and we expect that rate-independent hysteretic energy dissipation can also be found in other types of soft condensed matter, including biological materials.  

\section*{Author contributions}
\textbf{Robert Magerle:} Conceptualization (equal); Funding acquisition (equal); Methodology (equal); Resources (lead); Supervision (lead); Formal analysis (supporting); Visualization (supporting); Writing – original draft (equal); Writing – review \& editing (equal).
\textbf{Paul Zech:} Data Curation (supporting); Formal analysis (lead); Investigation (equal); Methodology (equal); Software (equal); Visualization (supporting); Writing – original draft (equal); Writing – review \& editing (equal).
\textbf{Martin Dehnert:} Data Curation (lead); Formal analysis (supporting); Investigation (equal); Methodology (equal); Software (equal); Visualization (lead); Writing – original draft (equal); Writing – review \& editing (equal).
\textbf{Alexandra Bendixen:} Conceptualization (equal); Funding acquisition (equal); Supervision (supporting); Writing – original draft (supporting); Writing – review \& editing (supporting).
\textbf{Andreas Otto:} Conceptualization (equal); Methodology (equal);  Writing – original draft (supporting); Writing – review \& editing (supporting).

\section*{Conflicts of interest}
The authors declare no competing interests.

\section*{Acknowledgements}
We acknowledge funding from the Volkswagen Foundation (grants I/77476 and 93794) and the Deutsche Forschungsgemeinschaft (INST 270/152-1 FUGG).
We thank G. Radons for discussions and N. Moore for proofreading. 

\section*{Appendix}
\subsection*{A\quad Materials and methods} 
\subsubsection*{A.1\quad Sample preparation}
Polished, $\qtyproduct{1 x 1}{\centi\metre}$ large Si substrates were cleaned in a 1:1 mixture of acetone/toluene and mounted on a heating plate ($\SI{150}{\degreeCelsius}$). Subsequently,  residual organic contamination was removed with a \ch{CO2} jet (Snowjet, Spraying Systems Co., USA).
A chicken's calcaneus (Achilles) tendon was removed from a fresh chicken leg sourced from a local supermarket. A piece of about $\SI{2}{\milli\metre}$ in length was extracted and spread out onto a polished Si substrate. After drying, the specimen was washed three times with deionized water, air-dried, and rehydrated with humid air in the AFM. The samples contain regions with bundles of collagen fibrils as well as isolated collagen fibrils.

\subsubsection*{A.2\quad Atomic force microscopy}
For AFM measurements, a NanoWizard II instrument (JPK Instruments AG, Berlin, Germany) was used in conjunction with two types of silicon cantilevers (see Table\,\ref{tab:table_cantilevers}). These stiff cantilevers minimize the jump-to-contact related hysteresis in the force--distance (FD) data.
Measurements on hydrated collagen fibrils were performed at room temperature with 88\%--90\% relative humidity (RH) using a lab-built setup similar to that used in ref. \citenum{Spitzner2015,Uhlig2017,Magerle2020}. The specimen was allowed to equilibrate for at least $\SI{40}{\minute}$ before AFM measurements were performed. 

\begin{table}[ht!]
\caption{\label{tab:table_cantilevers}
AFM cantilevers and their characteristic parameters.}
\begin{tabular}{p{3.0cm}p{2.6cm}p{2.3cm}} 
\hline
 &Type I&Type II\\ \hline 
Manufacturer& \mbox{NanoWorld AG} (Neuch\^{a}tel, Switzerland)& NANOSENSORS (Neuch\^{a}tel, Switzerland)\\
Model& NCH& PPP-NCSTR\\
Tip radius$^a$ & $<\SI{8}{\nano\metre}$& $<\SI{7}{\nano\metre}$\\
Half opening angle$^a$& 35°&  15°\\
Backside coating& none & Al\\
Length$^a$& $\SI{125}{\micro\metre}$& $\SI{150}{\micro\metre}$\\
Width$^a$& $\SI{30}{\micro\metre}$& $\SI{27}{\micro\metre}$\\
Thickness$^a$& $\SI{4}{\micro\metre}$& $\SI{2.6}{\micro\metre}$\\
Resonance frequency$^b$& $\SI{330}{\kilo\hertz}$, $\SI{354}{\kilo\hertz}\,$$^d$& $\SI{159}{\kilo\hertz}$\\
Quality factor$^b$& $439.4$, $476.6$$^4$ & $210.9$\\
Spring constant$^c$& $\SI{26.08}{\newton\per\metre}$, $\SI{23.18}{\newton\per\metre}\,^d$ & $\SI{4.67}{\newton\per\metre}$\\
\end{tabular} 
$^a$As specified by the manufacturer. $^b$Measured values. $^c$Determined as in ref. \citenum{Sader1999}. $^d$Cantilever used for measuring stress relaxation and creep.
\end{table}

\subsubsection*{A.3\quad Force--distance measurements}
In a first set of experiments (Series I), we studied an isolated collagen fibril deposited on a Si substrate using Type I cantilevers. 
We measured FD data sets with multiple intermediate retract-approach cycles at 10 positions with $\SI{100}{\nano\metre}$ pixel distance along the fibril's crest, with tip velocities $v_\text{tip}$ ranging from $\num{0.01}$ to $\SI{4}{\micro\metre\per\second}$. The resulting FD data are shown in Figs.\,\ref{fig:play_experiment}, \ref{fig:play_experiment_vel} and \ref{fig:creep_relaxation}.

In a second set of experiments (Series II), we studied a collagen fibril within a bundle of many fibrils using a Type II cantilever. For each tip velocity, we measured rows of 32 FD data sets with $\SI{32}{\nano\metre}$ pixel distance along the fibril's crest within a pristine part of the same collagen fibril (shown in Fig.\,\ref{fig:forc}a). The corresponding FD data are shown in Figs.\,\ref{fig:forc}, \ref{fig:Edis_experiment} and \ref{fig:fit}.

\subsection*{B\quad Complex contact stiffness} 
\begin{figure}[ht!]
    \centering
    \includegraphics[width=\columnwidth]{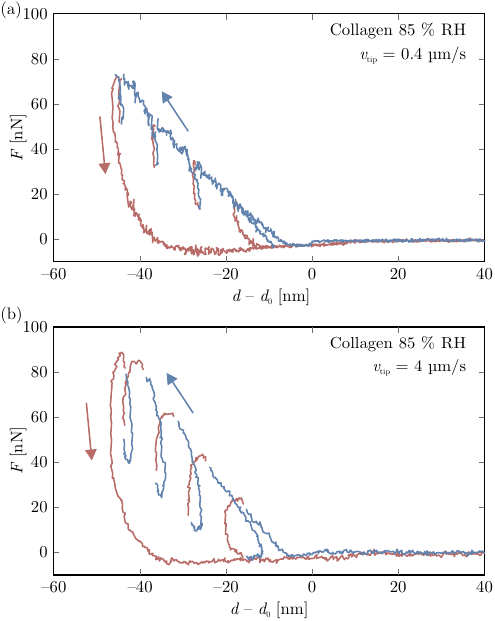}
    \caption{
    \begin{sansmath}
    Force trajectories with multiple retract-approach cycles measured on a hydrated collagen fibril using (a) $v_\text{tip} = \SI{0.4}{\micro\metre\per\second}$ and (b) $v_\text{tip} = \SI{4}{\micro\metre\per\second}$. The measurements are made on an isolated collagen fibril deposited on a Si substrate.
    \end{sansmath}
    }
    \label{fig:play_experiment_vel}
\end{figure}

For larger tip velocities, viscous friction causes the formation of inner hysteresis loops (Fig.\,\ref{fig:play_experiment_vel}).
This effect can be described with a complex contact stiffness $k_T^*(\omega)$ as a function of frequency $\omega$ corresponding to the collagen fibril's complex dynamic modulus.\cite{Grant2012}

\subsection*{C\quad Stress relaxation and creep}
\begin{figure}[ht!]
    \centering
    \includegraphics[width=\columnwidth]{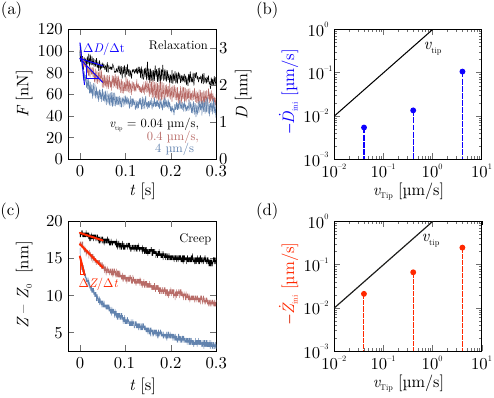}
    \caption{
    \begin{sansmath}
    Stress relaxation and creep are slower than tip velocity and do not affect FD data measured in cyclic indentation experiments. (a) Force relaxation $F(t)$ and the corresponding cantilever deflection $D(t)$ (solid lines) for different $v_{\text{tip}}$. (b) Initial deflection rate $-\dot{D}_\text{ini}$ as function of $v_{\text{tip}}$. (c) Creep $Z-Z_0$ for different $v_{\text{tip}}$. (d) Initial deflection rate $-\dot{Z}_\text{ini}$ as function of $v_{\text{tip}}$. The bisections $-\dot{D}_\text{ini}=v_{\text{tip}}$ and $-\dot{Z}_\text{ini}=v_{\text{tip}}$ are shown as black lines. The measurements are made on an isolated collagen fibril deposited on a Si substrate.
     \end{sansmath}
    }
    \label{fig:creep_relaxation}
\end{figure}

We measured stress relaxation and creep at the crest of isolated collagen fibrils deposited on a Si substrate at 85\% RH using Type I cantilevers. 
For the stress relaxation measurement, the tip is made to indent with $v_{\text{tip}}$ until a repulsive force of $\SI{100}{\nano\newton}$ is reached. 
Then, the tip position is held constant, at which time the evolution of $F(t)$ is measured via the corresponding cantilever deflection $D(t)$ (Fig.\,\ref{fig:creep_relaxation}a). 

For measuring creep, the tip is indented until a repulsive force of $\SI{100}{\nano\newton}$ is reached. 
After this, the force is kept constant, and the tip indentation $Z-Z_0$ is monitored as a function of time (Fig.\,\ref{fig:creep_relaxation}c). 
We obtain the initial deflection relaxation rate $\dot{D}_\text{ini} = [D(\Delta t)-D(0)]/ \Delta t $ from the slope of $D(t)$ for $t \in [0, \Delta t]$, where $t = 0$ is the end of the tip approach period and $\Delta t$ is indicated by the length of the slope shown as blue lines in Fig.\,\ref{fig:creep_relaxation}a. 
The initial deflection relaxation rate $\dot{D}_\text{ini}$ and the initial creep rate $\dot{Z}_\text{ini}$ (determined analogously to $\dot{D}_\text{ini}$) are the upper boundaries for the relaxation rates (creep rates) during the stress relaxation (creep) experiment, as indicated with the dashed lines in Fig.\,\ref{fig:creep_relaxation}b, d. 
Both stress relaxation and creep, proceed more slowly than the tip velocity and, therefore, do not influence the time evolution of $F(t)$ occurring during the measurement of FD data. 

\balance


\bibliography{rsc} 
\bibliographystyle{rsc} 

\begin{acronym}
\acro{FD}[FD]{force--distance}
\acro{AFM}[AFM]{atomic force microscopy}
\acro{RPM}[RPM]{return point memory}
\end{acronym}

\end{document}